\documentclass[fleqn,twoside]{article}
\usepackage{espcrc2}

\usepackage{graphicx}
\usepackage[figuresright]{rotating}

\usepackage{lineno}

\newcommand{\AmS}{{\protect\the\textfont2
  A\kern-.1667em\lower.5ex\hbox{M}\kern-.125emS}}

\title{Upper Limit on the  Diffuse Flux of Cosmic  $\nu_\mu$ with the ANTARES Neutrino Telescope}

\author{Simone Biagi\address{Dipartimento di Fisica dell'Universit\`a and INFN -- Sezione di Bologna, \\
		Viale Berti Pichat 6/2, 40127 Bologna, Italy}\thanks{simone.biagi@bo.infn.it}
		for the ANTARES collaboration}
             
\begin{document}

\begin{abstract}

A search for a diffuse flux of astrophysical muon neutrinos, using data collected by the ANTARES neutrino telescope from December 2007 to December 2009 is presented. A $(0.83\times 2\pi)$ sr sky was monitored for a total of 334 days of equivalent live time.
The searched signal corresponds  to  an excess of events,  produced  by astrophysical sources, over the expected atmospheric neutrino background without any particular assumption on the source  direction.  
Since the number of detected events is compatible with the number of expected background events, a 90\% c.l. upper limit on the diffuse $\nu_\mu$ flux with a $E^{-2}$ spectrum is set at
$E^2\Phi_{90\%}  =   5.3 \times 10^{-8}   \  \mathrm{GeV\ cm^{-2}\ s^{-1}\ sr^{-1}}$ 
in the energy range  20 TeV -- 2.5 PeV. 
Other signal models with different energy shape were also tested and some rejected.  
\end{abstract}

\maketitle

\section{Introduction}\label{sec:introduction}

The ANTARES high-energy neutrino telescope is a three-dimensional array of photomultiplier tubes  (PMT) distributed over 12 lines installed deep in the Mediterranean Sea, each line including 75 PMTs \cite{Paschal}. 
A neutrino telescope in the Northern hemisphere includes the Galactic Centre in its field of view and is complementary to the IceCube Antarctic telescope \cite{Teresa}.

The main goal of the experiment is the search for high-energy neutrinos from astrophysical sources. 
If the sensitivity of point source search techniques is too small to detect neutrino fluxes from individual sources, it is possible that many sources could produce an excess of events over the expected atmospheric neutrino background. 
In this proceeding the search for very-high energy extraterrestrial muon neutrinos from unresolved sources is presented using data collected by the ANTARES telescope from December 2007 to December 2009.

Atmospheric muons  and neutrinos  are the main sources  of background in a neutrino telescope.  The former  can be suppressed by applying requirements on the  direction  of the events, 
the latter is an irreducible background. 
As the spectrum of cosmic neutrinos is expected to be harder than that of atmospheric neutrinos,  the signal we are looking for    corresponds  to  an excess of high energy events   in the 
measured energy spectrum without any particular assumption on the source  direction.

Electrons (in the so-called ``leptonic models'') or protons and nuclei (``hadronic models'') can be accelerated in astrophysical processes.
Hadronic models   \cite{chiarusi} predict that   the energy produced  in  the sources is carried away by cosmic rays, $\gamma$-rays and neutrinos. 
A benchmark flux  for the measurement of diffuse neutrinos is the Waxman-Bahcall (W\&B) upper bound \cite{wb}. 
Using  the CR observations at $E_{CR}\sim 10^{19}$ eV ($E_{CR}^2 \Phi_{CR} \sim 10^{-8}$ GeV cm$^{-2}$s$^{-1}$sr$^{-1}$) the diffuse flux of muon  neutrinos is constrained at the value:
\begin{equation}
E^2_\nu \Phi_\nu < 4.5/2 \times 10^{-8} \ \mathrm{GeV\  cm^{-2}\ sr^{-1}\ s^{-1}}
\label{eq:wb}
\end{equation}
(the factor 1/2 is added to take into account neutrino oscillations).


\section{Neutrino tracking and energy reconstruction}\label{sec:tracking}

Muon neutrinos are detected via charged current interactions: $\nu_\mu + N \rightarrow \mu+X$. 
The arrival times and the amplitudes of the Cherenkov light signals      detected by the PMTs  \cite{antaresDAQ}  are used to reconstruct the trajectory of  muon neutrinos  and to estimate their energy.  

The track reconstruction algorithm defined in \cite{aart_icrc09} is based on a likelihood fit that uses a detailed parametrization of the probability density function for the photon arrival times. 
The track position and direction, the information on the number of hits ($N_{hit}$) used for the reconstruction and a quality parameter $\Lambda$ are the main outputs of the algorithm.  $\Lambda$ is determined from the likelihood and the number of compatible solutions found by the algorithm itself.  $\Lambda$  can be used to reject badly reconstructed events.
For $E_\nu>$ 10 TeV,  an angular resolution for muon neutrinos better than 0.3$^\circ$  is accomplished  by the ANTARES detector.


\subsection{Monte Carlo simulations}\label{sec:MC}

The Monte Carlo (MC) simulation tools \cite{brunner,mar5line} include  the production of Cherenkov light, the generation of the optical background  caused by  radioactive isotopes  and bioluminescence present in sea water, and the    digitization of the PMT signals. 
In particular, the PMT simulation also includes the probability of a detected hit giving rise to an afterpulse.  The simulation of afterpulses  is critical  when the energy estimator  defined in \S \ref{sec:R} is applied to MC events.
The afterpulse  probability was measured  in laboratory using ANTARES \cite{anta_nim} and NEMO \cite{nemo_nim}  PMTs and it was confirmed with  deep-sea data.
Upgoing  muon neutrinos and downgoing atmospheric muons have been simulated and stored in the same format used for the data.

\textbf{Signal and atmospheric neutrinos.}
MC muon neutrino events have been generated in the  energy range  $10 \leq E_\nu \leq 10^8$  GeV and zenith angle between $0^\circ \leq \theta \leq 90^\circ $ (upgoing events). 
The same MC sample can be differently weighted to reproduce the ``conventional'' atmospheric neutrino spectrum   (Bartol),  $\propto E_\nu^{-3.7}$ at high energies  \cite{bartol}, and  the expected  astrophysical signal spectrum, $\propto E_\nu^{-2}$. 
The normalization of the signal flux  is irrelevant when defining cuts,   optimizing procedures  and calculating  the sensitivity. Here a diffuse flux test signal is defined equal to:
\begin{equation}
E_\nu^2\ \Phi_{\nu} =  1.0 \times 10^{-7} \  \mathrm{GeV\ cm^{-2}\ s^{-1}\ sr^{-1}}.
\label{eq:test_limit}
\end{equation}

Above 10 TeV,  the semi-leptonic decay of short-lived charmed particles $D \rightarrow K + \mu + \nu_\mu$ becomes a significant source of atmospheric ``prompt leptons''.
The Recombination Quark Parton Model (RQPM)  is used in this simulation, since it  gives the largest prompt contribution  among    the models considered  in \cite{prompt_Costa}.


\textbf{Atmospheric muons.}
The ANTARES trigger rate  \cite{anta_line1}   is dominated by atmospheric muons that represent the main background for a neutrino telescope. A small fraction (approximately 5\%) of triggered downgoing muons is mis-reconstructed as upgoing; their rejection is a crucial point in this analysis.

The MUPAGE package \cite{mupage} was used to simulate atmospheric muon samples. 
One year of equivalent live time with a total energy $E_T\geq$ 1 TeV and bundle multiplicity $m = 1 \div 1000$ was generated.  The total energy $E_T$ is the sum of the energy of the individual muons in an atmospheric muon bundle. 
Triggered ANTARES  events mainly consist of multiple muons  originating in the same primary CR interaction \cite{antares_low}.


\subsection{Energy dependent variable}\label{sec:R}

\begin{figure}[!tb]
\includegraphics[width=78mm]{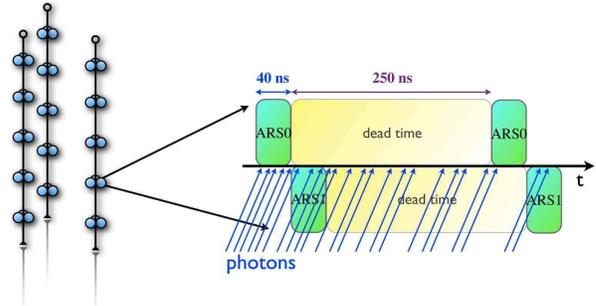}
\vskip -.8cm
\caption{\small{Definition of the variable $R_i$ on the $i$-th PMT. In this example, both ARS0 and ARS1 are fired; after the integration dead-time, both chips collected light again. In this example,  $R_i=4$.}}
\vskip -.4cm
\label{fig:Repetitions}
\end{figure}

The only way to separate atmospheric and astrophysical neutrinos is through a discrimination based on the energy.
An original energy estimator is defined, which is  based on hit repetitions  in the PMTs due to the different arrival time of \textit{direct} and \textit{delayed} photons (Fig. \ref{fig:Repetitions}). 
\textit{Direct} photons are emitted at the Cherenkov angle and  arrive at the PMTs without being scattered.
Radiative processes contribute to  energy losses  linearly with the muon energy for $E_\mu> 1$ TeV.
The resulting electromagnetic showers produce additional light.  
Photons originating from secondary electromagnetic showers  or scattered Cherenkov radiation  arrive on the PMTs \textit{delayed} with respect to the \textit{direct} photons, with arrival time differences up to hundreds of ns \cite{chiarusi}. 
The fraction of \textit{delayed} photons increases with the muon energy.



The signal  produced by  the PMTs  is processed by two   Analogue Ring Sampler (ARS)  \cite{ars_paper}
 which digitize the time and the amplitude of the signal (the $hit$). They are operated in a token ring scheme. If the signal crosses a preset threshold, typically 0.3 photo-electrons, the first ARS integrates the pulse within a window of 40 ns. 
If triggered, the second chip provides a second hit with a further integration window of 40 ns. After digitization, each chip has a dead time of typically 250 ns. After this dead time, a third and fourth hit can also be present.

\begin{figure}[!tb]
\includegraphics[width=75mm]{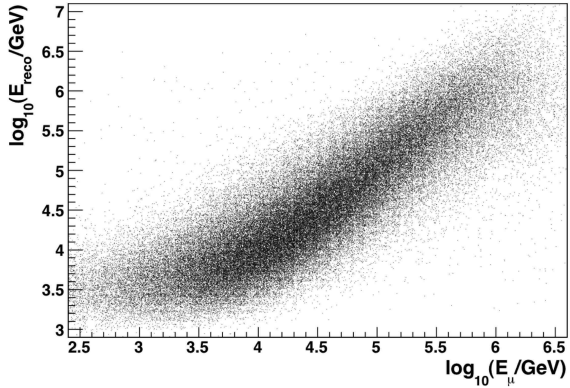}
\vskip -.6cm
\caption{\small{Reconstructed muon energy using the $R$ energy estimator  as a function of the true muon energy.  
The displayed  events are selected after the quality cuts defined in \S \ref{sec:muon_rejection}. Only neutrino-induced muons are selected.  }}
\vskip -.4cm
\label{fig:R_vs_Emu}
\end{figure}

The number of repetitions $R_i$ for the $i$-th PMT is defined as the number of hits in the same PMT within 500 ns from the earliest hit selected by the reconstruction algorithm (Fig. \ref{fig:Repetitions}). In most cases, $R_i$ =1 or 2, but it could be also 3 or 4. The mean number of repetitions in the event is defined as $R = \frac{\sum R_i}{N_{PMT}} $, where $N_{PMT}$ is the number of PMTs in which hits selected  by the tracking algorithm are present.
For a muon neutrino sample, $R$ is linearly correlated with the log of the true muon energy $E_{true}$ in the range from 10 TeV ($\overline R \simeq 1.26$) to 1 PeV ($\overline R \simeq 1.73$). 
$R$ can be used to  estimate  the muon energy  $E_{reco}$, see Fig. \ref{fig:R_vs_Emu}. The distribution of log$(E_{reco}/E_{true})$ has a FWHM=0.8.  The  resolution is comparable or better with respect to other energy reconstruction algorithm \cite{est_zornoza}.

This  energy estimator is robust because it does not depend on the number of active PMTs and on non-linear effects on charge integration.

\section{Cosmic neutrino signal selection}

The data collected  from December 2007 to December 2009 are analyzed.
In this period, the detector configuration changed several times with 9, 10 and 12 active lines. For this reason, three different detector configurations, based on the number of active lines and optical modules, were reproduced  in MC simulations. 
Data  runs are selected according to the  data-quality  requirements   explained in  \cite{mar5line}.
The total live time is 334 days:  70 days with 12 lines, 128 days with 10 lines and 136 days with 9 lines.

\subsection{Rejection of atmospheric muons} \label{sec:muon_rejection}

\begin{table}[tb]
\caption{{\small  Number of expected events  for  MC and data.}}
\label{tab:prelim} 
\small \begin{tabular}{ccccc}
\hline
      &  $\mu_{\mathrm{atm}}$ & $\nu_{\mathrm{atm}}$ & $\nu_{\mathrm{sig}}$ & Data\\ 
\hline
Reco &  $2.2 \cdot10^8$ & $7.1 \cdot 10^3$ & 106 & $2.5 \cdot10^8$  \\ 
Upgoing &$4.8 \cdot10^6$ & $5.5 \cdot 10^3$ & 80 & $5.2 \cdot10^6$  \\ 
1$^{\mathrm{st}}$-level & $9.1 \cdot10^3$ & 142 & 24 & $1.0 \cdot10^4$  \\ 
2$^{\mathrm{nd}}$-level & 0 & 116 & 20 & -- \\
\hline
\end{tabular}
{\small Expected events in 334 days of equivalent live time for the three MC samples: atmospheric muons, atmospheric neutrinos (Bartol+RQPM), astrophysical signal from eq. \ref{eq:test_limit},  and data. Reco: at the reconstruction level; Upgoing:  reconstructed as upgoing; 1$^{\mathrm{st}}$-level: after the first-level  cuts; 2$^{\mathrm{nd}}$-level: after the second-level cut.}
\vskip -.4cm
\end{table}

\begin{figure*}[tb]
\includegraphics[width=77mm]{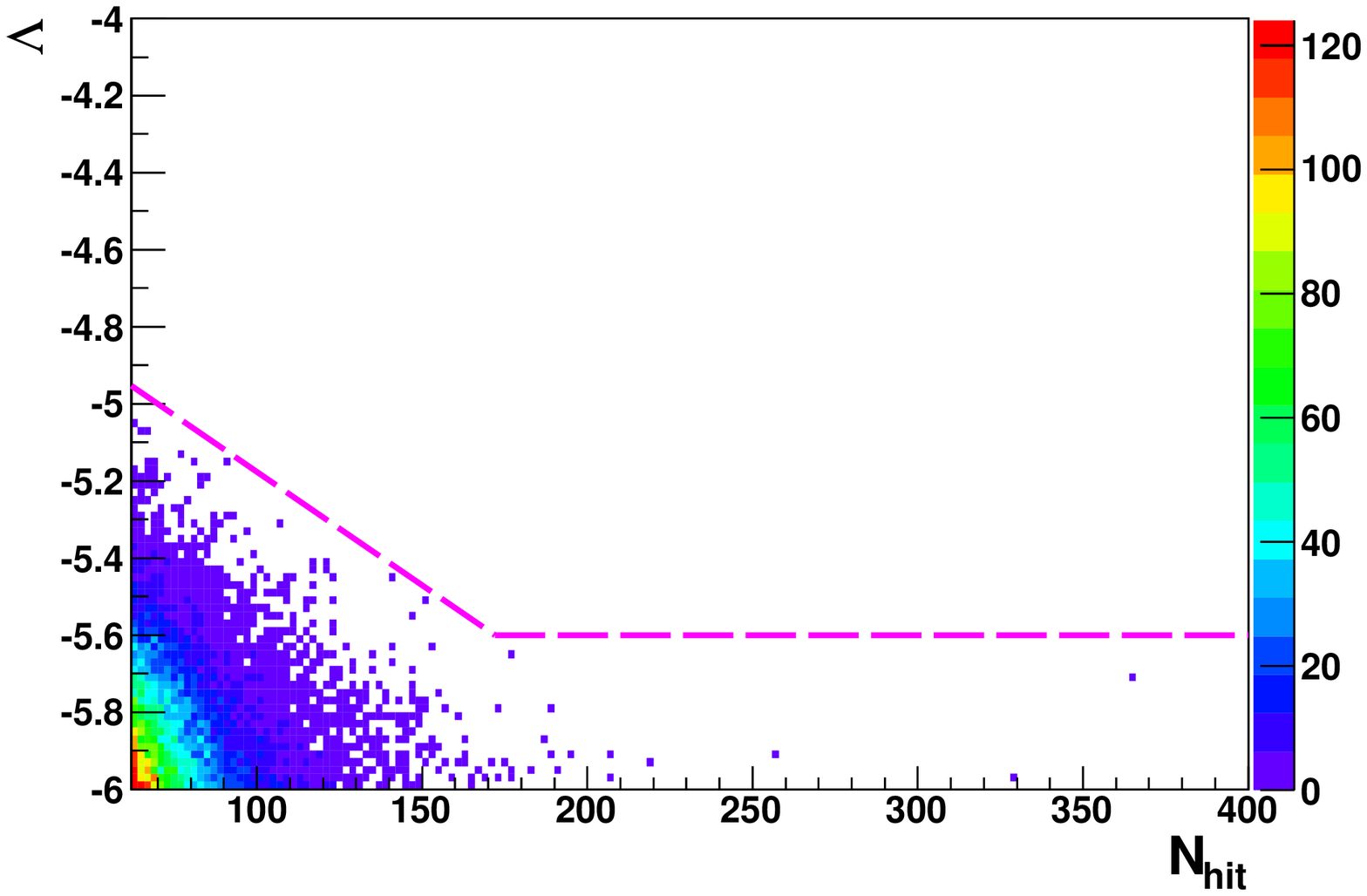}
\ \ \ \ \ \ \ \ 
\includegraphics[width=77mm]{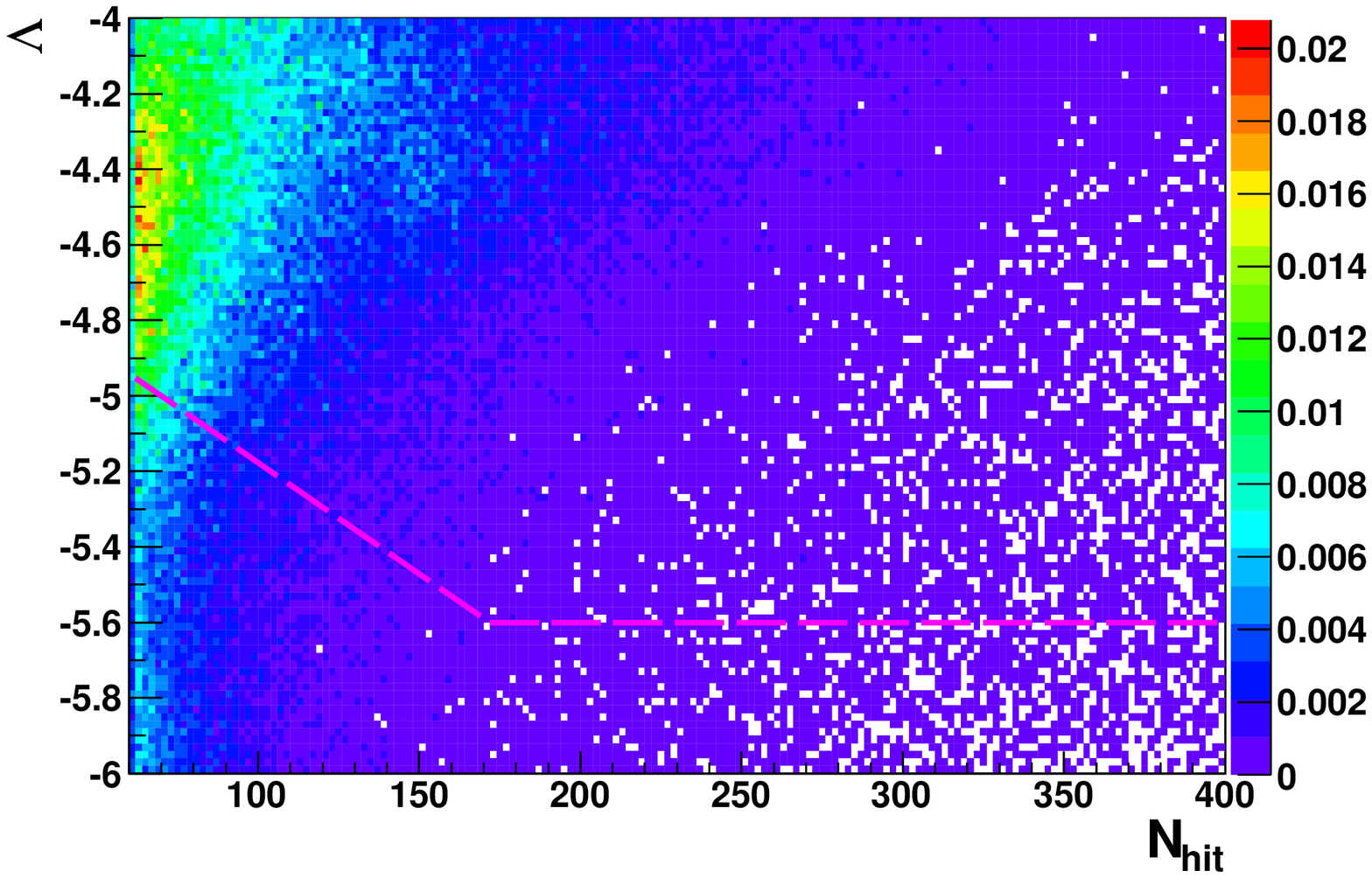}
\vskip -1.cm
\caption{\small{Scatter plots of the reconstruction quality parameter  $\Lambda$ versus the number of hits $N_{hit}$  for atmospheric muon (left) and   signal  neutrino (right) MC simulations   after the first-level  cuts. The pink line  represents the second-level cut described by eq.  \ref{eq:intercut}. The color code is in units year$^{-1}$.}}
\vskip -.2cm
\label{fig:second_level_cut}
\end{figure*}

As described  in \S \ref{sec:MC}, the rejection of atmospheric muons is a crucial point in the search for a cosmic neutrino signal.
This contamination can be strongly suppressed  by applying requirements on the geometry of the events and on the track reconstruction quality parameter $\Lambda$.
Two different levels  of cuts  are defined in order  to remove the contamination of mis-reconstructed  atmospheric muons  from the final sample.

\noindent \textbf{First-level cuts.} Events are selected according to these criteria:
$(i)$ upgoing particles with reconstructed zenith angle $\theta_{rec} < 80^\circ$ (corresponding to 0.83$\times 2\pi$ sr); 
$(ii)$ ${\Lambda > -6}$; $(iii)$ ${N_{hit} > 60}$;
$(iv)$ reconstruction with at least two lines. 
The first-level cuts     reduce   the rate of mis-reconstructed events by almost 3 orders of magnitude, as indicated in Table \ref{tab:prelim}.

\noindent \textbf{Second-level cut.} 
The remaining  atmospheric muons have the  quality parameter $\Lambda$ which on average decreases with increasing $N_{hit}$.  Fig. \ref{fig:second_level_cut} (left) shows the correlation between  $\Lambda$ and $N_{hit}$  for atmospheric muons. 
In order to completely remove  the expected rate of mis-reconstructed events in the MC sample, a cut value  
$\Lambda^*$ is defined as a function of $N_{hit}$: 
\begin{equation}
\Lambda^* = \left\{ \begin{array}{ll}-4.59-5.88 \cdot 10^{-3} N_{hit} \ \ &  N_{hit} \leq 172  \\
	 -5.60 &   N_{hit} >   172    \end{array}\right  .
\label{eq:intercut}
\end{equation}
Removing all events with $\Lambda<\Lambda^*$,  the atmospheric muons are completely suppressed.
Independent MC atmospheric muon simulations using CORSIKA (see details in \cite{mar5line}) confirm that the maximum contamination in the final sample is less than 1 event/year. 
As can be seen in Fig. \ref{fig:second_level_cut} (right), the signal  is highly   preserved from the second-level cut.   The effects of the first- and second-level cuts on signal and atmospheric neutrinos  are also given in Table \ref{tab:prelim}.



\subsection{Discrimination from atmospheric neutrinos}\label{sec:MRF}

A cut on  the energy dependent variable $R$, defined in \S \ref{sec:R},  is used to separate   the diffuse flux signal from the atmospheric $\nu_\mu$ background.
The  optimal cut value is obtained through a blinding procedure on MC events, without using informations from the data. 
The numbers of expected   signal ($n_s$) and  background ($n_b$) events  are computed as a function of $R$.  
Then, calculating   the so-called Model Rejection Factor (MRF) defined in \cite{mrp},  the best cut is obtained and 
used as the discriminator between \textit{low energy} events, dominated by the atmospheric neutrinos, and \textit{high energy} events, where the signal could exceed  the background. 
After the optimization of all the parameters,   the   observed data events ($n_{obs}$) are revealed (\textit{un-blinding procedure}) and compared with the expected background  for the selected region of $R$. 
If data are  compatible with the background, the upper limit  for the signal flux  is calculated using the Feldman-Cousins method \cite{feldman} at a 90\% confidence level (c.l.).



The cumulative distributions of the $R$ variable  for atmospheric neutrino background  (Bartol+RQPM) and diffuse flux signal (eq. \ref{eq:test_limit}) are computed   for the three  configurations of the ANTARES detector and the corresponding live times. 
The  MRF  is calculated as a function of  $R$ using these cumulative distributions. 
The minimum   
  found for $R=1.31$ determines the cut value for the energy dependent variable  \cite{papero_diffusi}. 
Assuming  the Bartol (Bartol+RQPM) atmospheric $\nu_\mu$ flux, 8.7 (10.7) background events and  10.8 signal events  are expected for $R\ge 1.31$.
The central 90\% of the signal is found  in the neutrino energy range   $20 \  \mathrm{TeV} < E_{\nu} < 2.5 \  \mathrm{PeV}$.


\section{Upgoing neutrino candidates}\label{sec:unblind}

\begin{figure}[!tb]
\includegraphics[width=75mm]{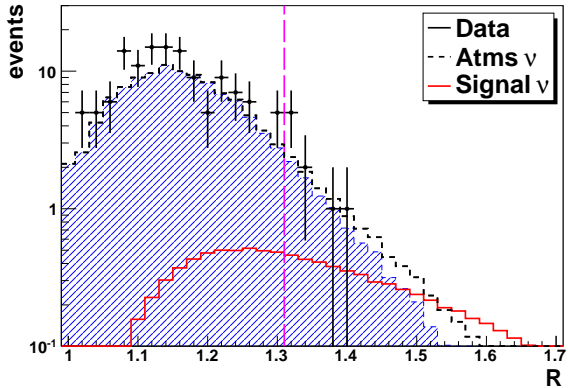}
\vskip -.6cm
\caption{\small{Distribution of the $R$ parameter for the 134 neutrino candidates in the 334 days of equivalent live time. Points represent data, the filled histogram is the atmospheric neutrino MC (Bartol model only). The dashed line represents the maximum contribution (RQPM) of  ``prompt'' neutrinos. The MC predictions are not normalized to the data. The signal normalized at  the upper limit (eq. \ref{eq:upper_limit}) is shown as a full line. The cut at $R=1.31$ is indicated as a vertical line.}}
\vskip -.4cm
\label{fig:neutrino_unblinded}
\end{figure}

\subsection{Low energy events   $R<1.31$}

Events surviving the second-level cut are upgoing neutrino candidates.
The first step of the un-blinding is to reveal the events with $R<1.31$. In this region, 125 events are found. A comparison  with MC predictions  is shown in  Fig.  \ref{fig:neutrino_unblinded}  as a function of $R$.
The events with $R\ge 1.31$ in Fig. \ref{fig:neutrino_unblinded} are   uncovered    after the final un-blinding of the data sample.
MC predictions are lower by  $\sim$ 20\%  with respect to the detected events.
Bartol  atmospheric neutrino MC predicts   104.0 events  with  $R<1.31$, and 
Bartol + RQPM   predicts 105.2 events. The discrepancies 
between predicted and measured events 
are  well within the systematic uncertainties of the absolute neutrino flux at these energies (25-30\%) \cite{bartol}.

The number of expected background events with $R\geq1.31$  is 8.7  for Bartol MC only.
Most prompt models described in \cite{prompt_Costa} give negligible contributions; the RQPM model  predicts the largest contribution of 2.0 additional events with respect to the conventional Bartol flux.
An average over all the considered models gives a contribution of 0.3 events.
A combined model of Bartol flux plus the average contribution from prompt models  is adjusted with the data/MC normalization factor  obtained in the   $R<1.31$ region. Hence  the number of expected background events for $R\ge 1.31$  is  10.7.

\subsection{High energy events and upper limit}

The MC simulations have been tested and compared with data.  In particular, the $R$ distributions show a reasonable agreement  both for atmospheric muons \cite{papero_diffusi} and for atmospheric neutrinos in the low energy  region $R<$1.31 (c.f. Fig. \ref{fig:neutrino_unblinded}).  
As a consequence, the  signal  region with  $R\ge1.31$ was un-blinded and  9 high-energy neutrino candidates are found.

Systematic uncertainties on the expected number of background events in the $R\ge 1.31$  region are evaluated considering: 
$(i)$ the contribution of prompt neutrinos, estimated as ${}^{+1.7}_{-0.3}$ events;  $(ii)$ the uncertainties from the conventional neutrino flux, that depend mainly on the uncertainty on the absolute flux as a function of the energy and on the spectral index, evaluated  to  be     $\pm 1.1$ events. 
The uncertainties on the detector efficiency (angular acceptance of the optical module, water absorption, scattering length, trigger simulation and the effect of   afterpulses) amount to 5\%:   they affect the detection both of  signal and background neutrinos in the high energy region.  

\begin{figure}[!tb]
\center{\includegraphics[width=75mm]{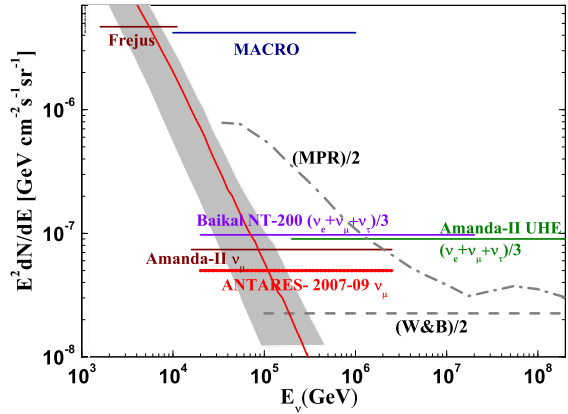}}
\vskip -.6cm
\caption{\small{
The ANTARES 90\% c.l. upper limit  for a $E^{-2}$ diffuse  $\nu_\mu+\overline \nu_\mu$ flux obtained with this analysis, compared with the results obtained by   other experiments and theoretical predictions.
See \cite{chiarusi} and references therein.
}}
\vskip -.4cm
\label{fig:upper_limits}
\end{figure}

The number of observed events is compatible with the number of expected background events. 
The $90\%$  c.l. upper limit on the number of signal events $\mu_{90\%}(n_ {obs}, n_b)$ for $n_{obs}=9$ observed events and $n_b =10.7\pm 2$ background events including the systematic uncertainties is computed with the method of \cite{conra}:   the value $\mu_{90\%}(n_b) = 5.7$ is obtained. 
The   upper limit  on the diffuse flux  is given by $\Phi_{90\%} = \Phi_\nu \cdot \mu_{90\%}/ n_s$:
\begin{equation}
E^2 \Phi _{90\%}  =   5.3 \times 10^{-8}   \  \mathrm{GeV\ cm^{-2}\ s^{-1}\ sr^{-1}} .
\label{eq:upper_limit}
\end{equation}
This limit holds for the energy range
between  20 TeV to 2.5 PeV.
The result is compared with other measured flux upper limits  and theoretical predictions  in Fig. \ref{fig:upper_limits}.

Some theoretical  predictions  of  cosmic neutrino fluxes with a spectral shape different from $E^{-2}$ are also tested.
For each model a cut value $R^*$ is optimized following the procedure in \S \ref{sec:MRF}. Table \ref{tab:models} gives the  results for the models tested: the values of  $R^*$, the numbers $N_{mod}$ of $\nu_\mu$ signal events for $R\ge R^*$, the energy intervals where 90\% of the signal is expected, the ratios between $\mu_{90\%}$ (computed according to \cite{feldman}) and $N_{mod}$. A value of  $\mu_{90\%}/N_{mod}<1$ indicates that the theoretical model is inconsistent with the experimental result at the 90\% c.l. \cite{papero_diffusi}.

\section{Conclusions}

Using   data from 334 days of equivalent live time  collected   with   the ANTARES   telescope,     a search for a diffuse flux of high energy cosmic muon neutrinos  was made.    
The  90\% c.l. upper limit on the diffuse $\nu_\mu$ flux with a $E^{-2}$ spectrum is set at
$E^2\Phi_{90\%}  =   5.3 \times 10^{-8}   \  \mathrm{GeV\ cm^{-2}\ s^{-1}\ sr^{-1}} $ in the energy range  20 TeV -- 2.5 PeV. Other signal models with different energy shape are also tested and some of them excluded at the 90\% c.l..

\begin{table}[tb]
\caption{{\small  Tested flux models.}}
\label{tab:models} 
\small \begin{tabular}{ccccc}
\hline
Model   & R$^*$ & N$_{mod}$ & $\Delta E_{90\%}$ & $\mu_{90\%}/N_{mod}$ \\ 
& & &(PeV)  & \\
\hline
MPR				& 1.43 & 3.0 & 0.1$\div$10 & 0.4 \\ 
P96$p\gamma$	& 1.43 & 6.0 & 0.2$\div$10 & 0.2 \\ 
S05				& 1.45 & 1.3 & 0.3$\div$ 5 & 1.2 \\ 
SeSi				& 1.48 & 2.7 & 0.3$\div$20 & 0.6 \\ 
M$pp+p\gamma$	& 1.48 & 0.24 & 0.8$\div$50 & 6.8 \\ 
\hline
\end{tabular}
{\small Astrophysical flux models, the value of the R$^*$ which minimizes the MRF, the expected number of events $N_{mod}$, the energy range $\Delta E_{90\%}$ in which the 90\% of events are expected, and the ratio $\mu_{90\%}/N_{mod}$.  See \cite{papero_diffusi} and references therein. }
\end{table}

\end{document}